\documentclass[usenatbib]{mn2e}
\usepackage{graphicx,color,epsfig}
\newcommand{\apj}{ApJ}

\newcommand{\mnras}{MNRAS}
\newcommand{\icarus}{ICARUS}
\newcommand{\aap}{A\&A}
\newcommand{\araa}{ARA\&A}
\newcommand{\apjl}{ApJL}

\def\ltsima{$\; \buildrel < \over \sim \;$}
\def\simlt{\lower.5ex\hbox{\ltsima}}
\def\gtsima{$\; \buildrel > \over \sim \;$}
\def\simgt{\lower.5ex\hbox{\gtsima}}

\newcommand\mj{{\,{\rm M}_{\rm J}}}

\def\del#1{{}}
\title[Coreless giant planets]{Positive metallicity correlation for coreless giant planets}

\author[S. Nayakshin]{Sergei Nayakshin\\ 
Department of Physics \& Astronomy,
  University of Leicester, Leicester, LE1 7RH, UK\\
{E-mail:~} {\rm Sergei.Nayakshin@le.ac.uk}}

\begin{document}

\date{Received}

\pagerange{\pageref{firstpage}--\pageref{lastpage}} \pubyear{2008}

\maketitle

\label{firstpage}

\begin{abstract}
Frequency of detected giant planets is observed to increase rapidly with
metallicity of the host star. This is usually interpreted as evidence in
support of the Core Accretion (CA) theory, which assembles giant planets as a
result of formation of a massive solid core. A strong positive
planet-metallicity correlation for giant planets formed in the framework of
Gravitational disc Instability (GI) model is found here. The key
novelty of this work  is ``pebble accretion'' onto GI fragments which has
been recently demonstrated to accelerate contraction of GI fragments. Driven
closer to the star by the inward migration, only the fragments that accrete
metals rapidly enough collapse and survive the otherwise imminent tidal
disruption. The survival fraction of simulated planets correlates strongly
with the metallicity of the host star, as observed.
\end{abstract}

%\begin{keywords}{}
%\end{keywords}

\section{Introduction}\label{sec:intro}

In its original form, GI hypothesis posited that gas giants are made in
  situ by gravitational fragmentation of a massive protoplanetary disc
  \citep{Kuiper51,Boss97}. This has been correctly criticised
  \citep{Rafikov05} since the discs can actually fragment only beyond $\sim$
  tens of AU \citep{Rice05}. However, since simulations show that GI fragments
  migrate in rapidly \citep{VB06,BaruteauEtal11}, it appears perfectly
  feasible for them to form at $\sim 100$~AU but then end up arbitrarily close
  to the parent star due to disc migration. Furthermore, some of the fragments
  could give birth to Earth-mass or more massive cores due to grain
  sedimentation \citep[e.g.,][]{Boss97,HS08,Nayakshin10b}.  If these gas
    fragments are tidally disrupted, only the cores survive
  \citep{BoleyEtal10}, which potentially provides a new pathway to forming
all kinds of planets at all separations in a single framework that was called
``Tidal Downsizing'' \citep[TD;][]{Nayakshin10c}.

However, there appears to be a major inconsistency of TD/GI with
observations. Giant planets are much more frequent around metal-rich hosts
\citep{Gonzalez99,FischerValenti05}.  Radiative contraction of GI planets is slower at
high metallicities \citep{HB11}, hence predicting fewer planets surviving
tidal disruption. Core Accretion \citep[CA; e.g.,][]{PollackEtal96} paradigm
is, in contrast, consistent with the metallicity trend and explains it as a
consequence of a more robust massive core assembly at high metallicities.

Nayakshin (2014b) showed that accretion of medium sized grains
\citep[``pebbles'', ][]{JohansenLacerda10} onto pre-collapse gas fragments
actually speeds up their contraction and collapse. In this picture giant
planets collapse not due to emission of radiation (like stars do) but due to
accretion of metals in small grains, which acts as an effective cooling
mechanism (cf. equation \ref{etot} below and figure 1). 
This paper presents first detailed coupled planet-disc evolutionary calculations of TD
hypothesis that incorporate this new physics, treating non-linear disc-planet
interaction, the rate of grain deposition into the planet and its response to
that in detail. A grid of models covering a reasonable range in poorly
constrained parameters of the model (such as grain opacity, disc viscosity,
etc.)  is calculated to delineate statistical trends of the model. A
strong positive correlation of planet survival probability with metallicity of
the host is found.

We also note in passing that Bowler et al (2015, arXiv:1411.3722) find that
giant gas planets are extremely rare at large $\sim 100$~AU separation from
their parent stars, which the authors interpret as evidence that gravitational
instability does {\em not} produce giant planets often. This interpetation of
the data is based on outdated ideas in which GI planets do not migrate. Modern
simulations
\citep[e.g.,][]{BoleyEtal10,BaruteauEtal11,ChaNayakshin11a,ZhuEtal12a} all
show that that GI clumps migrate in rapidly.  Another interpretation of the
Bowler et al (2015) results, consistent with the papers cited above and the
calculations below, is that most of GI fragments migrate closer in to the star
and are either tidally destroyed and became terrestrial like planets or
survive the disruption and became hot jupiters instead.

\section{Numerical Methods}

In the protoplanetary disc environment, both gas and grains are
gravitationally attracted to massive bodies embedded in it, but gas has
pressure gradient forces able to resist the pull, whereas grains do
not. Grains that are moderately weakly coupled to gas via aerodynamical
friction -- grains of a few cm in size, $a_{\rm peb}$, in the inner disc, but $1$ mm or
less in the outer disc -- are captured by the body most efficiently
\citep{JohansenLacerda10,OrmelKlahr10}. Pebble accretion rate appropriate for
the massive planets that we study here is $\dot M_z = 2 R_H^2 v_K \Sigma_p/a$
\citep[e.g.,][]{LambrechtsJ12}, where $v_K$ is Keplerian velocity at the
planet's location, $a$; $\Sigma_p = f_p z_d \Sigma_d$ and $\Sigma_d$ are the
surface densities of pebbles and gas, respectively, and $0 \le f_p < 1$ is the
fraction of pebbles in the total grain surface density ($z_d \Sigma_d$). We
assume that $f_p$ increases linearly with $z_d$ due to a more rapid grain
growth at higher $z_d$, so $f_p = f_{\rm p0} (z_d/z_\odot)$, where $f_{\rm
  p0}=$~const$\ll 1$ is a free parameter, and $z_d$ and $z_\odot=0.015$ is the
disc and solar metallicities, respectively.

The disc surface density at the planet's location, $\Sigma_d$, is not
independent of the planet, as the planet interacts with the disc strongly.
Following \cite{NayakshinLodato12}, the protoplanetary disc is described by a
viscous azimuthally symmetric time-dependent model that includes the tidal
torque of the planet on the disc
\begin{equation}
  \frac{\partial\Sigma}{\partial
  t}  = \frac{3}{R}\frac{\partial}{\partial R}
  \left[R^{1/2}\frac{\partial}{\partial R}(R^{1/2}\nu\Sigma)\right]
  -\frac{1}{R}\frac{\partial}{\partial
  R}\left(2\Omega R^2\lambda\Sigma\right) 
\label{eq:diffplanet}
\end{equation}
where $\Sigma$ is the disc surface density at radius $R$, $\lambda$ is the
tidal torque from the planet. The torque can be either in type I (no gap) or
type II (a gap in the disc is opened). Two-dimensional hydro simulations
\citep{CridaEtal06} show that a deep gap in the disc is opened when parameter
${\cal P} = 3H/4R_H + 50 \alpha_{SS} (H/a)^2 (M_*/M_p)\simlt 1$, where $H$ is
the disc vertical scale height at $a$, and $\alpha_{ss} \ll 1$ is the disc
viscosity parameter. Based on this result, we smoothly join the type I and
type II regimes using the self-consistently found value of ${\cal P}$ at the
location around the planet. This approach is necessary since migration
  rates of planets depend sensitively on whether a gap in the disc is opened
  or not \citep{GalvagniMayer14}.

Nayakshin (2014b), \cite{Nayakshin14b}, uses a 1D spherically symmetric
radiative hydrodynamics (RHD) code with grains treated as a second fluid to
simulate contraction of an otherwise isolated planet that accretes
grains. Such an approach is unfortunately too computationally expensive in the
framework of a full disc-planet interaction problem, and forced previous
workers to use analytical models for the planets
\citep[e.g.,][]{Nayakshin10c,ForganRice13b}.

Here, a simpler "follow the adiabat" approach
\citep{MarleauCumming14,FortneyHubbard04} in which the planet is assumed
isentropic, is used. This is a reasonable approximation since the energy transfer
inside the planet is strongly dominated by convection even at Solar
opacities/metallicities \citep{HB11}. For a given initial conditions, e.g.,
the planet mass, $M_p$, the central fragment's temperature, $T_c$, and grain
properties for each radial zone in the planet, a solution of the equilibrium
equations is found by iterations on the central gas density. This determines
planetary radius, $R_p$, and the total energy of the planet, $E_{\rm tot}$,
which is then evolved in time according to
\begin{equation}
{d E_{\rm tot}\over dt} = - L_{\rm rad} - {G M_p \dot M_z\over R_p}\;,
\label{etot}
\end{equation}
where $L_{\rm rad}$ is the radiative luminosity of the planet, and the last
term on the right is the change in the gravitational potential energy of the
planet due to grain accretion on it at the rate $\dot M_z$. After evolving
$E_{\rm tot}$ by a small amount, a grain growth step of same duration then
follows. The new total energy and grain properties in every zone in the planet
then allow us to determine the new planet's structure, which is found by
iterating on both $T_c$ and the central gas density. The procedure is then
repeated. We tested the isentropic approach against the RHD code for a number
of fragment contraction cases, including metal loading tests on the planet,
and found an acceptable (typically $\sim 10-20$\%) agreement.

It is important to point out the following. Pebbles sedimenting down onto
  the planet do so at differential velocities (usually linearly proportional
  to their size, $a_{\rm peb}$). From experiments it is well known that grains
  colliding at velocities exceeding a few m~s$^{-1}$ fragment
  \citep[e.g.,][]{BlumWurm08}. In the context of TD, \cite{Nayakshin14b}
  showed that grain-grain collisions limits the grain size inside the planets
  to a few cm. Therefore, the velocity with which pebbles impact the planet,
  $v_{\rm imp}$, cannot be much larger than a few m~s$^{-1}$, which is very
  small compared to the escape velocity from the planet, $\sqrt{2 G
    M_p/R_p}\approx 1500$~m~s$^{-1}$, where $M_p = 1 \mj$ and $R_p = 1$~AU is
  used. For this reason the kinetic energy input term in equation \ref{etot},
  $\dot M_z v_{\rm imp}^2/2$, is neglected. This is in stark contrast to CA
  theory \citep[CA; e.g.,][]{PollackEtal96} where the solids from the disc
  enter the planet as planetesimals -- huge rocks very poorly coupled to gas
  -- and impact the growing planets at $v_{\rm imp}\simgt 2 G M_p/R_p$ and
  therefore heat the gaseous envelope strongly. Solids may nevertheless heat
  the gas envelope in both scenarios if they eventually reach a {\it massive}
  solid core in the planet. This effect is not considered here but is included
  in a follow-up paper.

Following \cite{HB11}, dust opacity in the fragment is directly proportional
to the metallicity of the gas, $z$: $ \kappa(\rho, T) = f_{\rm op}
\kappa_0(\rho, T) (z / z_\odot)$ where $\kappa_0(\rho, T)$ are the
interstellar gas plus dust opacities from \cite{ZhuEtal09} which assume Solar
metallicity, $z_\odot$, and $f_{\rm op}=$~const~$\le 1$ is a positive constant
which may be smaller than unity due to grain growth.

\section{Planet survival experiments}

\begin{figure}
\centerline{\psfig{file=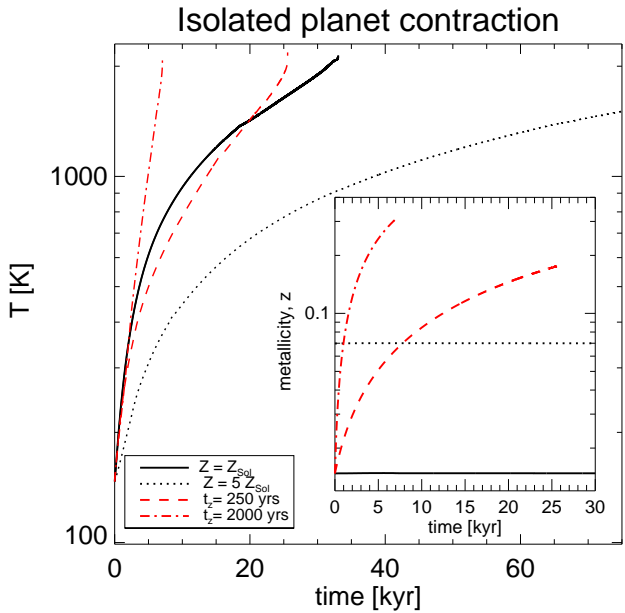,width=0.5\textwidth,angle=0}}
\caption{Radiative contraction of four GI fragments of mass $M_p=1\mj$. Black
  curves show $z=$~const planets, while red curves show two pebble accretion
  cases. The faster the metals are added to the planet, the faster it
  collapses to higher densities, increasing the chances of its survival.}
\label{fig:1}
\end{figure}

Before presenting the more complex planet-disc calculations, Figure
\ref{fig:1} shows evolution of central temperature for four isolated GI
fragments of mass $M_p=1\mj$.  Black curves are for two different constant
fragment metallicities, $Z=1$ and 5, where $Z=z/z_\odot$. The red curves are
for fragments with metal abundance increasing at rate 
\begin{equation}
\dot M_z  =  {z_\odot M_p \over t_{\rm z}}\;,
\label{tz}
\end{equation}
with $t_z$ labelled on the figure. The inset shows metallicity $z$ for the
four cases. The red and the $Z=1$ black curves end when $T_c$ reaches $\sim
2000$~K, at which point H$_2$ molecules dissociate, and the fragment collapses
dynamically to much higher densities. This marks formation of a dense young
Jupiter that could survive tides in the inner disc. The $Z=5$ fragment
contracts the slowest due to a high dust opacity. In contrast, $\dot M_z > 0$
fragments (red curves) contract faster than the $Z=1$ one. The metallicity of
the fragment corresponding to the dot-dashed curve is $Z=20$ yet it collapses
$\sim 5$ times faster than the $Z=1$ case. Such a collapse could be
  termed "dark" as relatively little radiation is emitted during contraction
  of the planet.

As explained in Nayakshin (2014b) in detail, and can also be qualitatively
seen from equation \ref{etot}, accretion of pebbles by the fragments is a form
of non-luminous cooling which is directly proportional to $\dot M_z$. This must
help fragments to survive. For example, if the fragments from figure 1 were
migrating inward on a time scale of $10^4$ years, the $Z=5$ one would have been
tidally destroyed in about 20,000 years, whereas the $t_z = 250$~years fragment
would have collapsed and could therefore continue to migrate almost
arbitrarily close to the star.

To quantify these ideas, Figure \ref{fig:disc_planet} tests survival of
fragments born at $a_0=120$~AU for the pebble accretion model at three disc
metallicities, $Z_d \equiv (z_d/z_\odot)= 0.5,$ 1 and 2 for the dashed, dotted
and solid curves, respectively. Colours are used to delineate different
quantities for the same fragment. The top panel shows time evolution of the
planet's separation, $a$, planet's radius, $R_p$, and Hill radius, $R_H = a
(M_p/3 M_*)^{1/3}$. Panel (b) shows metallicity of the planets, (c) shows
central temperature, $T_{\rm c}$, and (d) compares the migration time scale,
$t_{\rm migr} = - a/(da/dt)$, where $a$ is planet's semi-major axis, with the
metal loading time scale, $t_{\rm z}$, calculated 
  self-consistently. Namely, first $\dot M_z$ in the Hill's regime
  \citep[see][]{LambrechtsJ12} is found from the simulation, and then equation
  \ref{tz} is inverted to find $t_{\rm z}$.

Initially, $R_p\ll R_H$, so that tidal forces from the star are weak compared
to the planet's self-gravity. As the planet migrates closer in, $R_p$ and
$R_H$ decrease at different rates, and the planet is tidally disrupted if
$R_p\ge R_H$.

In all three cases, radiative cooling of the planet is negligible, that is,
$L_{\rm rad}$ is much smaller \citep[or even {\em negative} due to planet
  irradiation from the disc, see][]{VazanHelled12} than the last term in
equation \ref{etot}. The planets therefore contract mainly due to
accretion of pebbles. The higher the disc metallicity, the quicker the
planet's metallicity increases with time, and the faster it contracts (note
that $R_p$ decreases and $T_c$ increases). The lowest metallicity planet is
disrupted the soonest, at $a=3.3$~AU. $Z_d=1$ planet is about twice as
compact, so it makes it to $a=1.7$~AU before being disrupted. This planet
almost manages to collapse (reaches $T_c\approx 1500$~K), but $\dot M_z$
plummets when a deep gap around the planet is opened after $t\approx 55,000$
years (note that $t_z \rightarrow \infty$ at later times). Starved of
  metals, the planet stops contracting and gets disrupted soon thereafter. In
contrast, the $Z_d = 2$ planet contracts much more rapidly, and collapses at
$t\approx 33,000$ years, before it is tidally compromised. This planet could
be driven into the ``hot Jupiter'' region by a continuing disc migration, not
simulated here.

\section{A grid of models}

\begin{figure}
\centerline{\psfig{file=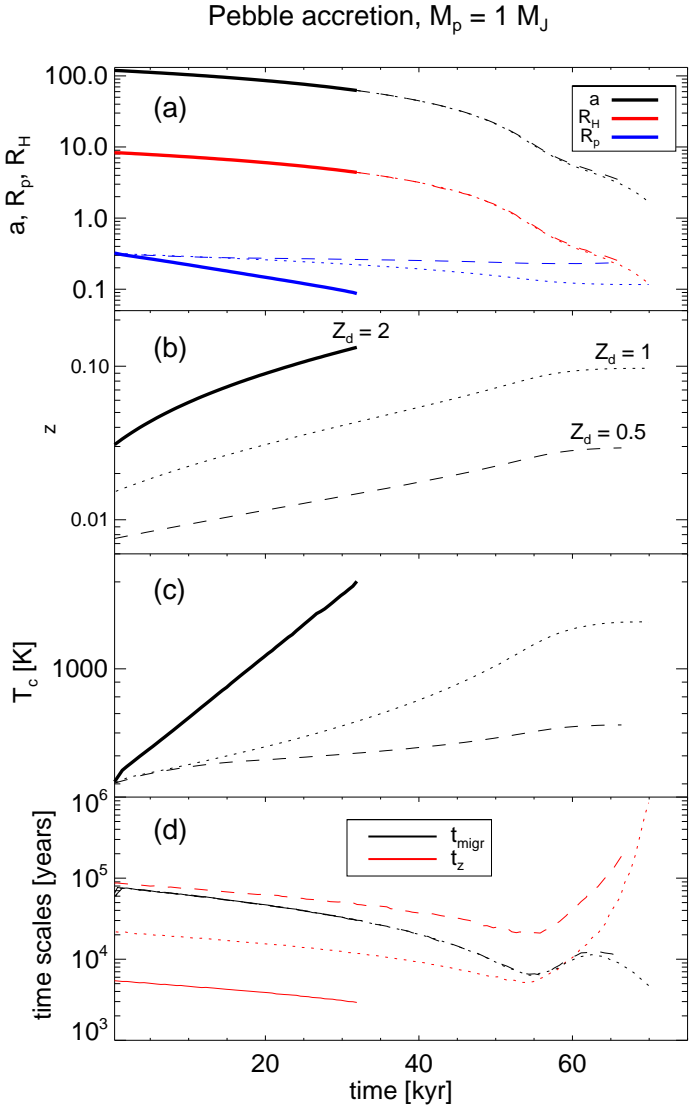,width=0.55\textwidth,angle=0}}
\caption{Evolution of a gas fragment accreting grains from the disc at three
  different disc metallicities ($Z_d=2$, 1 and 0.5, for solid, dotted and
  dashed curves, respectively).  Panels show: (a) planet-star separation, $a$,
  Hill's and planet's radii; (b) planet's metallicity, $z$; (c) central
  temperature of the planet; (d) migration and grain loading time scales. For
  all three cases, the fragment mass is $1\mj$, birth location $a=120$~AU,
  disc mass $100 \mj$ within 150 AU, viscosity parameter $\alpha_{\rm SS} = 0.02$,
  pebble fraction $f_{\rm p0}= 0.1$, and planet opacity parameter $f_{\rm
    op}=0.3$.  $Z_d=2$ fragment contracts rapidly, reaching $2000$~K and
  collapsing at $t\approx 32,000$ years. The metallicity of the planet is
  about $10 z_\odot$ at the point of collapse.  The $Z_d=1$ and 0.5 fragments
  contract less rapidly due to lower metal supply (resulting in longer $t_z$,
  see panel d), and are tidally disrupted at $a=1.7$ and 3.3~AU, respectively.
  The $Z_d=1$ planet would have actually collapsed if not for a deep gap in
  the disc, opened at $t\approx 55,000$ years, which cuts off grain accretion
  to almost zero.}
\label{fig:disc_planet}
\end{figure}

These results suggest that GI planets may be more likely to survive at higher
$Z_d$. To ascertain metallicity trends of the model, given large uncertainties
in the input physics, a grid of fragment survival experiments just like those
described in figure \ref{fig:disc_planet}, but now repeated for parameters
varied over a reasonably broad range, is run. Parameter values in the grid
are: disc viscosity $\alpha_{\rm SS} = 0.01, 0.02, 0.04$; planet's birth
location, $a_0 = 70, 120$; pebble mass fraction $f_{\rm p0} = 0.05$, 0.1 and
0.2. We also tested type I migration torque at $0.5$, 1 and 2 times that from
\cite{BateEtal03}. The grid of models is calculated for 9 different disc
metallicity values between $Z_d=1/3$ and $Z_d=3$, and the fraction of planets
surviving (that is collapsing before being tidally disrupted) is then found
for each metallicity bin. This comprises 486 planet survival experiments in
total.

%The opacity parameter is set to $f_{\rm op} = 0.5$.

\begin{figure}
\centerline{\psfig{file=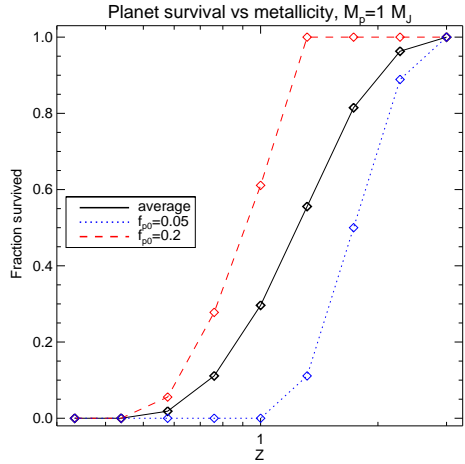,width=0.4\textwidth,angle=0}}
\caption{Planet survival probability versus $Z$, the metallicity in Solar
  units, for a planet of $M_p=1\mj$ mass and disc parameters covering a range
  of properties. The black diamonds show the full grid of models, while the
  blue and the red symbols show results for $f_{p0} = 0.05$ and $f_{p0}=0.2$
  (low and high pebble content, respectively). There is a strong positive
  planet survival correlation with the metallicity of the host disc.}
\label{fig:zcorr}
\end{figure}

\begin{figure}
\centerline{\psfig{file=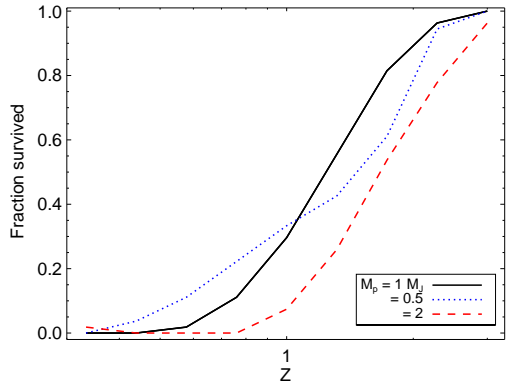,width=0.4\textwidth,angle=0}}
\caption{Planet survival probability versus $Z$ for three different planet's
  masses, as labelled on the figure. }
\label{fig:zmp}
\end{figure}

Figure \ref{fig:zcorr} shows the results for $M_{\rm p}=1\mj$. Fragments are
indeed much more likely to survive at high $z_d$ than they are at low
$z_d$. Black solid line shows the full grid of models, while the blue dotted
and the red dashed lines show $f_{\rm p0} = 0.05$ and 0.2 only,
respectively. The correlation with metallicity remains strong but is shifted
to higher $z_d$ (lower $z_d$) if pebbles are less (more) abundant. The
strength of the correlation appears sufficient to explain qualitatively the
observed trend of giant planet frequency being proportional to $Z^2$
\citep{FischerValenti05}. Figure \ref{fig:zmp} shows the same grid of survival experiments as shown in
fig. \ref{fig:zcorr}, but now conducted for different planet's masses. It is
seen that the correlation is qualitatively unchanged for $M_{\rm p} = 0.5\mj$
and $M_{\rm p}=2\mj$ planets \citep[which are abundant in
  the][sample]{FischerValenti05}.

The final mass of the planet may be different from the pre-collapse
   value for two reasons. Firstly, due to a substantial angular momentum of
   pre-collapse clumps \citep[e.g.,][]{BoleyEtal10,GalvagniEtal12}, not all of
   the planet's mass may end up in the planet, some may end up in the
   circum-planetary disc and then be lost. On the other hand, more gas could
   in principle be accreted from the disc onto the planet. These effects may
   extend the positive metallicity correlation found here for the $0.5-2\mj$
   planets to both lower and higher masses.

\section{Discussion and Conclusions}

Planet survival experiments in the context of Tidal Downsizing model for
planet formation \citep{Nayakshin10c} were performed. The new ingredient in
the model is ``pebble accretion'' of grains from the disc onto the fragments,
plus simultaneous treatment of the coupled planet and disc evolutionary
equations. Since pebble accretion accelerates collapse of gas fragments,
  a strong positive correlation of the fraction of survived giant planets
  versus metallicity of the host is found. TD/GI origin for giant gas planets
  is not, therefore, in conflict with the observed planet-metallicity
  correlation. Formation of solid cores within the planets is turned off in
  the present paper for simplicity but is to be considered and reported on in
  a forthcoming paper.

\cite{GalvagniMayer14} find results that are somewhat in disagreement with
ours. They find a much more copious production of giant planets without pebble
accretion which we found to be instrumental in driving the planets to collapse
here. This difference in the results may be in part due to a different
radiative cooling formalism used by \cite{GalvagniMayer14}. Namely, these
authors use earlier results of \cite{GalvagniEtal12} who studied disc
fragmentation and gas fragment collapse in 3D simulations, which is clearly
preferable to our 1D study in that aspect. On the other hand, in
\cite{GalvagniEtal12} the radiative cooling of the fragments is modelled with
a semi-analytical prescription \citep[as commonly done in 3D simulations of
  discs by a number of authors, see, e.g.][]{BoleyEtal10,ChaNayakshin11a}
rather than with a radiative transfer scheme. Our results, on the other hand,
are motivated (Nayakshin 2014b) by radiative hydrodynamics simulations of
contracting planets in which transfer of radiation is calculated with the
classical radiation diffusion approximation, albeit in 1D. Ideally, one would
like to combine 3D hydrodynamics with 3D radiative transfer to study formation
of giant planets. We must leave this challenging goal to future papers,
unfortunately.

\section*{Acknowledgments}

Theoretical astrophysics research in Leicester is supported by an STFC
grant. The author acknowledges useful comments on the manuscript by Richard
Alexander.  This work also used the DiRAC Complexity system, operated by the
University of Leicester, which forms part of the STFC DiRAC HPC Facility
(www.dirac.ac.uk). This equipment is funded by a BIS National E-Infrastructure
capital grant ST/K000373/1 and DiRAC Operations grant ST/K0003259/1. DiRAC is
part of the UK National E-Infrastructure.

%\bibliographystyle{mnras} 
%\bibliography{../nayakshin}

\label{lastpage}

\end{document}